\documentclass{jfm}

\usepackage{graphicx}
\usepackage{natbib}
\usepackage{amssymb,amsmath}
\usepackage{tabularx,booktabs}

\ifCUPmtlplainloaded \else
  \checkfont{eurm10}
  \iffontfound
    \IfFileExists{jfm-ifc/upmath.sty}
      {\typeout{^^JFound AMS Euler Roman fonts on the system,
                   using the 'upmath' package.^^J}%
       \usepackage{jfm-ifc/upmath}}
      {\typeout{^^JFound AMS Euler Roman fonts on the system, but you
                   dont seem to have the}%
       \typeout{'upmath' package installed. JFM.cls can take advantage
                 of these fonts,^^Jif you use 'upmath' package.^^J}%
      }
  \else
  \fi
\fi

\ifCUPmtlplainloaded \else
  \checkfont{msam10}
  \iffontfound
    \IfFileExists{amssymb.sty}
      {\typeout{^^JFound AMS Symbol fonts on the system, using the
                'amssymb' package.^^J}%
       \usepackage{amssymb}%
       \let\le=\leqslant  \let\leq=\leqslant
         
      }{}
  \fi
\fi

\ifCUPmtlplainloaded \else
  \IfFileExists{amsbsy.sty}
    {\typeout{^^JFound the 'amsbsy' package on the system, using it.^^J}%
     \usepackage{amsbsy}}
    {\providecommand\boldsymbol[1]{\mbox{\boldmath $##1$}}}
\fi

\newsavebox{\astrutbox}
\sbox{\astrutbox}{\rule[-5pt]{0pt}{20pt}}

\newcommand{\mb}{\mathbf}
\newcommand{\lbr}{\left(}
\newcommand{\rbr}{\right)}
\newcommand{\vl}{\vert}
\newcommand{\pd}{\partial}
\newcommand{\pdz}{\pd_z}
\newcommand{\pdn}{\nabla_{\bot}}
\newcommand{\Qn}{Q_{\bot}}
\newcommand{\Qz}{Q_{\parallel}}
\newcommand{\gammaz}{\gamma_{\parallel}}
\newcommand{\gamman}{\gamma_{\bot}}
\newcommand{\lsb}{\left[}
\newcommand{\rsb}{\right]}
\newcommand{\lb}{\left(}
\newcommand{\rb}{\right)}

\title{Fluid transport by individual microswimmers}

\author[D. O. Pushkin, H. Shum and J. M. Yeomans]%
{D\ls M\ls I\ls T\ls R\ls I\ns O.\ns P\ls U\ls S\ls H\ls K\ls I\ls N$^1$%
  \thanks{Email address for correspondence: mitya.pushkin@physics.ox.ac.uk},\ns
H\ls E\ls N\ls R\ls Y\ns S\ls H\ls U\ls M$^2$\break
\and J\ls U\ls L\ls I\ls A\ns M.\ns Y\ls E\ls O\ls M\ls A\ls N\ls S$^1$}

\affiliation{$^1$Rudolf Peierls Centre for Theoretical Physics, University of Oxford, 1 Keble Road, Oxford OX1 3NP, UK\\[\affilskip]
$^2$Department of Chemical and Petroleum Engineering, University of Pittsburgh, Pittsburgh, Pennsylvania, USA}

\pubyear{2012}
\volume{xx}
\pagerange{xxx--xxx}
\date{?; revised ?; accepted ?. - To be entered by editorial office}
\begin{document}

\maketitle

\begin{abstract}
We discuss the path of a tracer particle as a microswimmer moves past on an infinite straight trajectory. If the tracer is sufficiently far from the path of the swimmer it moves in a closed loop. As the initial distance between the tracer and the path of the swimmer $\rho$ decreases, the tracer is displaced a small distance backwards (relative to the direction of the swimmer velocity). For much smaller tracer-swimmer separations, however, the tracer displacement becomes positive and diverges as $\rho \to 0$. To quantify this behaviour we calculate the Darwin drift, the total volume swept out by a material sheet of tracers, initially perpendicular to the swimmer path, during the swimmer motion. We find that  the  drift can be written as the sum of a {\em universal} term which depends on the quadrupolar flow field of the swimmer, together with a non-universal contribution given by the sum of the volumes of the swimmer and its wake. The formula is compared to exact results for the squirmer model and to numerical calculations for a more realistic model swimmer.
\end{abstract}


\section{Introduction}

Microswimmers, by virtue of their size, typically have Reynolds numbers $\ll 1$. Hence they are well described by the Stokes equations. These are time independent and therefore, as first recognised by \citet{Purcell}, any swimming stroke that leads to motion at Re=0 must be non-invariant under time reversal. Swimming strategies evolved by microorganisms to overcome this restriction include waving flagella and rotating helical filaments.

Bacteria, algae and other microorganisms usually live in complex fluids, sharing their environment with a myriad of inert particles such as colloids and biofilaments. As they swim, the flow fields they produce stir the surrounding fluid, leading to the enhanced tracer diffusion observed in swimmer suspensions \citep{Wu00, Sokolov, Leptos, Kurtuldu, Mino11}, and predicted by simulations and using theoretical arguments \citep{Underhill, Rushkin, IshiPedley10, LinChildress, IZaid}. Bacterial stirring is thought to be an important factor in controlling and enhancing nutrient uptake.  Moreover, but still controversially, it has been suggested that stirring by small swimmers could provide a significant contribution to oceanic mixing \citep{KatijaDabiri, Visser}. 

\begin{figure}
\begin{center}
\includegraphics[width = 0.8\textwidth]{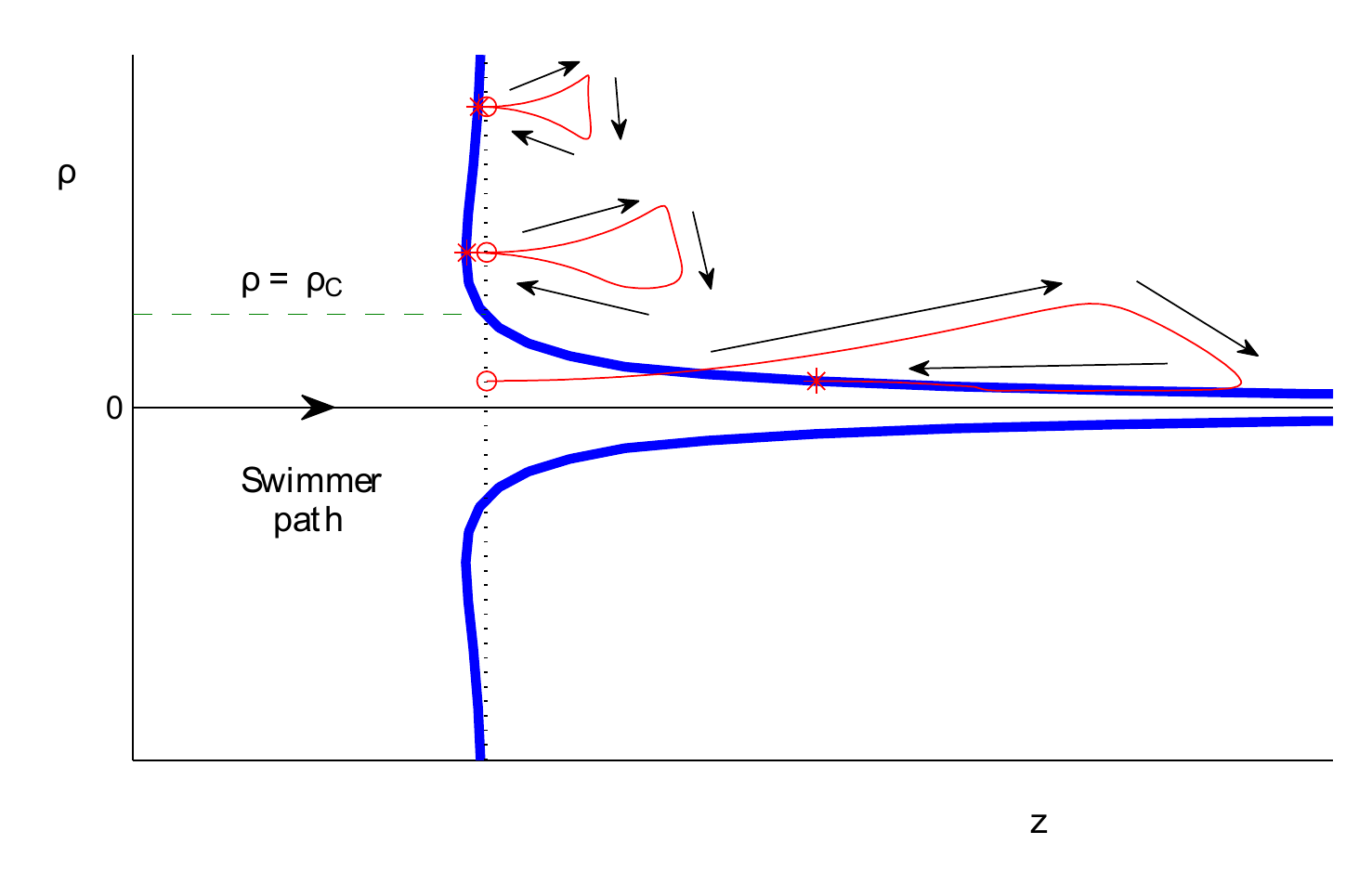}
\caption{Typical motion of a sheet of tracer particles as a swimmer moves in an infinite straight trajectory from left to right, along $\rho=0$  perpendicular to the sheet. The initial position of the tracer sheet is shown as a dotted line, and the envelope of the final tracer positions in blue. Examples of the loop like trajectories of the tracers are indicated as lighter, coloured lines. The starting and ending points of the tracer paths are marked by open circles and asterisks respectively and arrows indicate the direction of motion along the paths. For $\rho>\rho_C$ the net tracer displacement is negative and decays as $\rho^{-3}$ as $\rho \rightarrow \infty$ for dipolar swimmers. For  $\rho<\rho_C$ it is positive and diverges as $\rho \rightarrow 0$. However, the Darwin drift, the integrated volume swept out by the sheet, given by Eq~(\ref{eq:Darwin}), is finite.}
\label{fig:loops}
\end{center} 
\end{figure}

However, experimental trajectories of tracers in a suspension of swimmers often look loop-like \citep{Leptos}, with the net displacement between the initial and final locations considerably shorter than the characteristic trajectory size, Fig.~\ref{fig:loops}. Indeed, it has been shown that tracers advected by a passing swimmer follow closed trajectories when the swimmer, modelled as a viscous dipole or quadrupole, travels along a straight, infinite path \citep{Dunkel10}. This loop-like nature of the swimmer paths was found, on the average, to survive the smearing effect of Brownian fluctuations. 
	
Clearly, closedness of tracer trajectories severely limits the effectiveness of stirring, yet microswimmers enhance tracer diffusion. An explanation is called for, and this is an aim of this paper. We first, in Sec.~2, examine the tracer trajectories, identifying the properties of the flow field and the necessary assumptions and conditions that lead to loops. In particular we find that the trajectories are not closed for tracers near to the swimmer. Such particles are entrained by the swimmer, following it over a distance that diverges for a head-on collision. Loops are also destroyed by finite swimmer paths, and by the flow field appropriate to swimming in two dimensions.

In the remainder of the paper we focus on investigating entrainment in detail. To do this we calculate the Darwin drift, the volume of fluid moved by a swimmer as it travels along an infinite straight path \citep{Darwin, Benjamin, Eames94}. We find that although this quantity is infinite for a colloid pulled through a viscous fluid \citep{Eames03}, it is finite for a swimmer. Moreover it can be written as the sum of a universal term, which depends only on the quadrupole flow produced by the swimmer, and a non-universal contribution equal to the volume of fluid in the swimmer wake. 

The analytic development leading to this result is described in Sec.~3 where we consider the displacement of a material sheet of tracers lying perpendicular to the swimmer path. In Sec.~3a we calculate the far field tracer displacement showing that it scales as $\rho^{-3}$ for a dipolar swimmer, where $\rho$ is the initial distance between the tracer and the swimmer path. We then, in Sec.~3b, obtain an expression for the Darwin drift produced by a passing swimmer. Section 4 compares the theoretical results to calculations for two model swimmers: In Sec.~4a we present analytical results for the squirmer model.  Sec.~4b is devoted to a numerical calculation of the drift for a model bacterium which resembles {\em R. sphaeroides}.  Algebraic and numerical details appear in Appendices. Finally our results are summarised, and other mechanisms for tracer diffusion discussed, in Sec.~5.

\section{Geometrical constraints to mixing by swimmers}

Consider a single swimmer moving steadily in a quiescent fluid at a speed $V$ along the direction $\mathbf{k}$. Let $\mathbf{R}(t)$ is the time-dependent position of the swimmer and $\mathbf{u}\left( \mathbf{r},\mathbf{k} \right)$ the velocity induced at the position $\mathbf{r}$ relative to the swimmer. As the passive tracer moves with the local fluid velocity, its position $\mathbf{r}_T$ satisfies
\begin{equation}
\label{eq:tracer}
\frac{d \mathbf{r}_T}{dt} = \mathbf{u} \left(\mathbf{r}_T-\mathbf{R}(t),\mathbf{k} \right).
\end{equation}

A self-propelled swimmer must be force-free. This physical requirement imposes a strong restriction on its velocity field. In particular, the Stokeslet term of the multipolar expansion must vanish and the velocity far from the swimmer is generally dominated by the force dipole flow: 
 
\begin{equation}
\label{eq:dipole}
\mathbf{u}  \left(  \mathbf{r} ,\mathbf{k} \right) = - \kappa \left(  \mathbf{k}  \cdot \nabla \right) \mathbf{U}_S \left(  \mathbf{r} ,\mathbf{k} \right),
\end{equation}
where  $\mathbf{U}_S$ is the Stokeslet, the gradient operator $\nabla$ acts on the first argument of $\mathbf{U}_S$, and $\kappa$ is the dipole strength. The sign of  $\kappa$ distinguishes the far-field of pushers (such as {\it E. coli}), $\kappa>0$,  from that of the pullers (such as {\it C. reinhardtii}), $\kappa<0$. (Note that $\kappa$ needs to be multiplied by the prefactor $1/8 \pi \eta$, where $\eta$ is the kinematic fluid viscosity, in order to have units of a force dipole.) The Stokeslet can be expressed via the Oseen tensor $\mathbf{J}$: 
\begin{eqnarray}
\mathbf{U}_S  \left(  \mathbf{r} ,\mathbf{k} \right)= 
\mathbf{k}  \cdot  \mathbf{J}, \quad
\mathbf{J} =  \frac{\mathbf{E}}{r}  +  \frac{\mathbf{r}  \mathbf{r}}{r^3} . 
\end{eqnarray}
Here $r=\| \mathbf{r} \|$ and $\mathbf{E}$ is the identity matrix.  Thus, the dipolar flow field $\mathbf{u}$ decays at large distances as $r^{-2}$. 

Note that the expression (\ref{eq:dipole}) is considerably more restrictive than the general definition of the force dipole (also commonly referred to as stresslet and Stokeslet doublet) \citep{Chwang}. The strength of the general force dipole is a symmetric second order tensor and it may have six independent components. However, it is physically reasonable to expect that the lines of action of the resultant thrust and drag forces coincide and are aligned with the direction of swimming $\mathbf{k}$. These conditions reduce the number of independent components from six to one. 

Under this assumption the quadrupolar terms of the multipole expansion for an axisymmetric swimmer can be written
\begin{equation}
\mathbf{u^Q}  \lb  \mathbf{r} ,\mathbf{k} \rb = -\frac12 \lb \Qz (\mathbf{k} \cdot \nabla)^2 + \Qn \pdn^2 \rb \mathbf{U}_S  \left(  \mathbf{r} ,\mathbf{k} \right),
\label{e:quadrupole}
\end{equation}
where $\Qz$ and $\Qz$ are the force quadrupole strengths along and perpendicular to the swimming direction respectively and $\pdn^2$ is the Laplace operator in the plane normal to the swimming direction. 

In the particular case of $\Qz=\Qn=Q$,
\begin{equation}
\label{eq:quadsym}
\mathbf{u^Q}  \left(  \mathbf{r} ,\mathbf{k} \right) = -  \frac{Q}{2} \nabla^2 \mathbf{U}_S \left(  \mathbf{r} ,\mathbf{k} \right) \equiv Q \left(  \mathbf{k}  \cdot \nabla \right) \nabla  \left( \frac{1}{r} \right).
\end{equation}
A spherical body or a spherical droplet passively towed in viscous fluid has the quadrupole term of the same form \citep{Pozrikidis}. Remarkably, the flow field of a spherical particle moving in {\it inviscid} fluid also has the same form. Indeed, the expression to the right from the identity sign in (\ref{eq:quadsym}) is but a potential dipole \citep{Pozrikidis}. 

It is essential for our further argument that the quadrupole term (\ref{e:quadrupole}), similarly to the dipole term (\ref{eq:dipole}), can be expressed as a derivative along the swimming direction. Indeed,
\begin{align}
& \mathbf{u^Q}  \lb  \mathbf{r} ,\mathbf{k} \rb =  \left(  \mathbf{k}  \cdot \nabla \right) \mathbf{U^Q} \left(  \mathbf{r} ,\mathbf{k} \right), \label{e:quadgrad1} \\
& \mathbf{U^Q} \left(  \mathbf{r} ,\mathbf{k} \right) = \Qn \nabla \lb \frac{1}{r}  \rb + \frac{\Qn-\Qz}{2} 
\lb  \mathbf{k}  \cdot \nabla \rb 
\mathbf{U}_S  \left(  \mathbf{r} ,\mathbf{k} \right).
\label{e:quadgrad2}
\end{align}
Hence a general flow field produced by a swimmer takes the form
\begin{equation}
\mathbf{u}  \left(  \mathbf{r} ,\mathbf{k} \right) =  \left(  \mathbf{k}  \cdot \nabla \right)  \mathbf{U}_0 \left(  \mathbf{r} ,\mathbf{k} \right) + O(r^{-4}).
\label{eq:gradgen}
\end{equation}

We are now in a position to see that closedness of tracer trajectories is a universal feature rooted in the expression (\ref{eq:gradgen}).
Consider the velocity $\mathbf{U}_0( \mathbf{r}_T-\mathbf{R}(t))$ at the tracer location. Its Lagrangian derivative is
\begin{equation}
\frac{d \mathbf{U}_0}{dt}= \left( \mathbf{V} \cdot \nabla \right)  \mathbf{U}_0  -  \left(  \frac{d \mathbf{r}_T}{dt} \cdot \nabla \right) \mathbf{U}_0, \quad \mathbf{V}=V \mathbf{k}.
\end{equation} 
When the tracer velocity is negligible in comparison to the swimmer velocity, $\|  \frac{d \mathbf{r}_T}{dt}  \|  \ll  V $, the Lagrangian derivative coincides with the Eulerian derivative: 
\begin{equation}
\frac{d \mathbf{U}_0}{dt} \approx V \left( \mathbf{k} \cdot \nabla \right)  \mathbf{U}_0.
\end{equation} 
Combining this expression with Eqns.~(\ref{eq:tracer}) and (\ref{eq:gradgen}), results in:
\begin{equation}
\frac{d \mathbf{r}_T}{dt} \approx - \frac{\kappa}{V}  \frac{d \mathbf{U}_0}{dt}.
\end{equation}
Therefore, the overall displacement of the tracer as the swimmer travels along the infinite path equals:
\begin{equation}
\Delta \mathbf{r}_T= \int_{-\infty}^{+\infty}  \frac{d \mathbf{r}_T}{dt} dt = - \frac{\kappa}{V}  \left( \mathbf{U}_0(+\infty) -  \mathbf{U}_0(-\infty) \right) = \mathbf{0},
\end{equation}
i.e. the tracer trajectory is closed.

Besides translating, some swimmers, such as bacteria, often rotate around their bodies. But since typically the rotation axis coincides with the swimming direction, the induced azimuthal flow causes mixing only in the azimuthal direction. Our interest is in the cylindrically symmetrical transport and hence we have disregarded the azimuthal component of the flow. The validity of the representation (\ref{eq:dipole}) was verified in recent experiments for {\it E. coli} in \citep{Drescher}  and for {\it C. reinhardtii} in \citep{Guasto}. 

The assumptions inherent in proving that the tracer loops are closed can be summarised  as follows: \\
~\\
1) The (induced) tracer velocity is much smaller than the swimmer velocity.\\
2) The velocity field induced by the swimmer can be expressed in the form (\ref{eq:gradgen}). \\
3) The field $\mathbf{U}_0$ decays at infinity: $\mathbf{U}_0 (\mathbf{r}) \to 0$ as $r \to \infty$.\\
4) Swimmers move in straights paths, arriving from and leaving to infinity.\\

The effectiveness of biomixing \citep{Leptos, Kurtuldu} indicates that one or several of these assumptions are often violated in natural settings.  The character of such violations may serve as basis for systematic analysis of different mechanisms of biomixing, and we list examples in table \ref{Tab}.  In the remainder of this paper we provide a detailed analysis of situations where the first assumption is violated, and the tracer velocity is no longer small compared to that of the swimmer. This occurs for tracer particles that happen to lie in the vicinity of a swimmer path. Their motion can be understood as entrainment of the neighbouring fluid by a swimmer. Fuller comments on possible ways in which the other assumptions break down are given in the Discussion, Sec.~5.

\newcolumntype{Z}{>{\centering \arraybackslash}X}
\renewcommand {\tabularxcolumn}[1]{>{\arraybackslash}m{#1}}
\begin{table}
\caption{Necessary conditions for closed tracer loops and corresponding mixing mechanisms.}
\vskip 5mm
\label{Tab} 
\begin{tabularx}{\textwidth}{ZZZ}
\toprule 
{\bf Closed loops when} & {\bf Examples of violation }& {\bf Mixing mechanism }\\
\midrule 
{Induced tracer displacement $\ll$ Swimmer displacement}
 & {Tracers close to swimmer} & {Fluid entrainment} \\
\midrule 
\begin{equation*}
\mathbf{u}(\mathbf{r})=(\mathbf{k} \cdot \nabla) \mathbf{U_0}(\mathbf{r})
\end{equation*}
& Spinning bacterial cells  & Unimportant for axisymmetric swimmers \\
\midrule 
Short-range interactions: 
$ \| \mathbf{U_0}(\mathbf{r}) \|  \to 0\text{ as } r \to \infty$ 
& 2D dipole & Enhanced mixing in films\\
\midrule 
Swimmers arrive from and leave to infinity & Run-and-tumble of
{\it C.~reinhardtii} & 
Random re-orientations \citep{LinChildress} \\
\bottomrule 
\end{tabularx}
\end{table}

\section{Fluid entrainment by swimmers}

The fluid flow near a swimmer is determined by the swimmer shape and gait. Consequently, important physical quantities, such as the dissipated energy or the ability to stir the surrounding fluid may strongly depend upon the particular swimmer features. However, some aspects of the hydrodynamics, for example the dipolar far flow field, are {\it universal}, i.e. generic across a wide range of swimmers. When considering the effect of entrainment on particle transport we shall, in particular, identify the universal aspects of the tracer motion.



In order to analyse permanent displacement of fluid by a moving body it is convenient to envision an initially flat material sheet (ie a plane of tracer particles) of infinite extension, and a swimmer approaching it normally, see Fig.~\ref{fig:loops}. As the body moves, the sheet is deformed. Studies of the asymptotic shape of the material sheet and the fluid transfer fall under the rubric of the Darwin drift. The cases of towed bodies in inviscid \citep{Darwin, Benjamin, Eames94} and viscous flows \citep{Eames03} have been analysed in detail. Here we consider the Darwin drift for force-free swimmers, far from, and then closer to, the swimmer trajectory.

\subsection{Far field entrainment:}

The velocity field far from a swimmer is dominated by the dipolar and quadrupolar terms which, from Eqns.~(\ref{eq:dipole}) and (\ref{e:quadrupole}), are
\begin{align}
& \mathbf{u}  \left(  \mathbf{r} ,\mathbf{k} \right) = \lb - \kappa \pdz -\frac{\Qz}{2} \pdz^2 -\frac{\Qn}{2} \pdn^2 \rb \mathbf{U_S} \lbr  \mb r, \mb k \rbr + O  \left(  r^{-4} \right).
\label{eq:ufar}
\end{align}
Here we use Cartesian coordinates with the z-axis directed along the swimming direction and hence $(\mathbf{k} \cdot \nabla)=\pdz$.
The multipole expansion ensures that the asymptotic shape of the material sheet is naturally universal at large distances from the swimmer path. Let $\rho$ be the distance from this path to a tracer in the material sheet and $\Delta$ the resulting tracer displacement along the swimming direction, as the swimmer travels from and to infinity. Straightforward asymptotic calculations (see Appendix A) result in
\begin{align}
\label{eq:delta}
\Delta&=-C_1 \frac{\kappa^2}{V^2}\frac{1}{\rho^3}+C_2 \frac{\kappa \Qn}{V^2}\frac{1}{\rho^4} + O(\rho^{-5}),
&\text{for} \; 
\kappa \ne 0 . 
\end{align}
Here $C_1$ and $C_2$ are constants. Detailed calculation show that $C_1 = \pi/16$. A typical shape of the deformed sheet is shown in Fig.~\ref{fig:loops}. As $C_1>0$, far from the swimmer the sheet is displaced oppositely to the swimming direction both for pushers and pullers. Pushers and pullers are distinguished by the sign of the second term in (\ref{eq:delta}). Notably, the displacement $\Delta$ turns out independent of the quadrupole strength $\Qz$. The calculation we have just described is valid  in the region of small negative total tracer displacement. A quantitative comparison to numerical results will be given in Sec.~4b.

\subsection{Near field entrainment:}

Unlike the case of far distances, at close separations from the swimmer the shape of the deformed material sheet strongly depends on the swimmer size, shape, and stroke. It will, however, always remain integrable, and the volume of fluid entrained by the swimmer finite. We shall show, surprisingly, that the total volume of fluid displaced by the swimmer $v_D$ can be decomposed into a universal and a swimmer-dependent part having a lucid physical meaning. Before deriving this result, we pause to note that in his seminal paper \citet{Darwin} put forward a remarkable proposal that for a body moving steadily in an inviscid fluid, $v_D$ equals the added mass of the body. He derived this result for a sphere and a cylinder of circular shape, finding in accordance with his hypothesis $v_D=C_D v_0$, where $C_D=1/2$ for a sphere and $C_D=1$ for a cylinder. Later his arguments were refined  \citep[and references therein]{Eames94} and a notion of the partial Darwin drift was introduced to account for the case of a body starting a finite distance from the sheet before crossing it and moving to infinity. More recently, a spherical viscous drop towed through a fluid of different viscosity was considered \citep{Eames03}. It was found that in this case $v_D$ is divergent with a pre-factor proportional to the viscosity of the surrounding fluid. A viscosity-enhanced mechanism of ocean biomixing \citep{KatijaDabiri} was proposed based on this result. 

We follow the ideas of Darwin in our derivation of $v_D$ for force-free swimmers. We set the z-axis along the swimming direction and  
introduce the cylindrically symmetric stream function in the system of reference co-moving with the swimmer $\psi(\rho,z)$. For a swimmer with the velocity field (\ref{eq:ufar}) we obtain (see Appendix B): 
\begin{align}
\label{eq:streamf} 
\psi =  2 \pi \rho^2 & \lsb  \frac{V}{2} + \frac{\kappa z}{r^3} \right. \nonumber \\
 & \left.   - \lb 
\Qz \frac{1-3 \cos^2 \theta}{2} + \Qn \frac{1+3 \cos^2 \theta}{2} 
\rb 
\frac{1}{r^3} \rsb  + O(r^{-2}),
\end{align}
where $\quad r^2 =  \rho^2+z^2$ and $\quad \cos \theta = z/r$.

Next, we consider a flow region around the swimmer which has a finite lateral extension $-L_f \le z \le L_i$ and is enclosed in a streamtube of a finite initial radius $\rho_0$: $\{ \rho,z: \psi(\rho,z) \le \psi_0, \, -L_f \le z \le L_i, \, \psi_i=\psi(\rho_0,L_i)\}$. The total entrained volume is obtained from the volume entrained by the flow within this region by letting $\rho_0 \to \infty$. By a simple geometrical argument (see Appendix B and Fig.~\ref{DarwinPic}), the latter equals
\begin{equation}
v_D=A(\rho_0)-v_*. 
\end{equation}
Here $v_*$ is the volume of the flow region enclosed by a stagnation streamline around the swimmer (in the co-moving reference frame) and $A$ depends only on the far-field of the swimmer velocity:
\begin{align}
\label{eq:A}
A &= \frac{2 \pi \kappa}{V} \rho_0 \sin \theta \bigg|_{i}^{f} 
\nonumber \\
& + \frac{\pi (\Qz+\Qn)}{V} \cos \theta \bigg|_{i}^{f} 
+ \frac{\pi (\Qn -\Qz)}{V} \cos^3 \theta \bigg|_{i}^{f} + O(\rho_0^{-1}).
\end{align}
 At this point, however, Eq.~(\ref{eq:A})  is not well-defined, as the first term on the right hand side  may diverge as $\rho_0 \to \infty$. But it will remain finite, in fact equal to zero, if the swimmer path is symmetric with respect to the initial plane of the material sheet, $\theta_f=\pi-\theta_i$. Hence, under this condition, the dipole flow component does not contribute to the permanent volume transfer across the sheet and, when the swimmer travels an infinite distance,
\begin{equation}
A=\frac{2 \pi (\Qz+\Qn)}{V} + \frac{2 \pi (\Qn-\Qz)}{V}=\frac{4 \pi \Qn}{V}.
\label{eq:Af} 
\end{equation}

\begin{figure}
\begin{center}
\includegraphics[height=5cm]{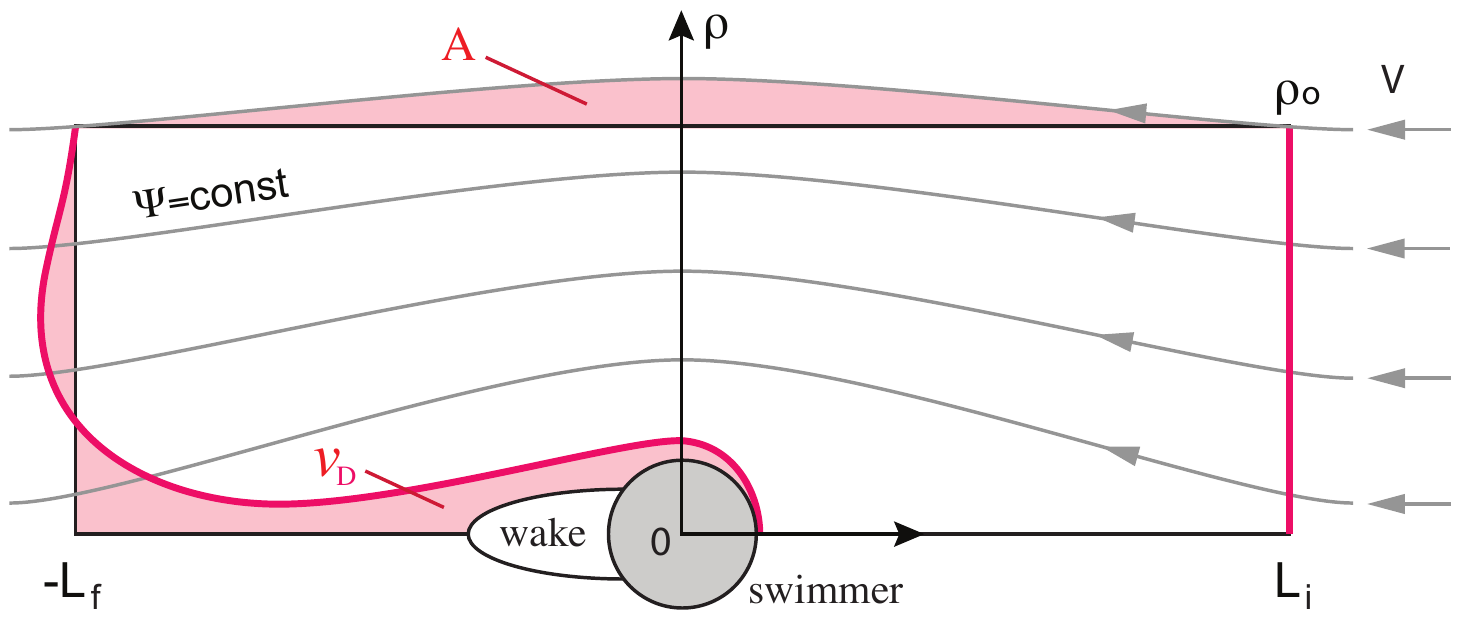}
\end{center}
\caption{Geometry and parameters used in the calculation of the Darwin drift in Appendix~B.}
\label{DarwinPic}
\end{figure}

The final expression for the Darwin drift reads
\begin{equation}
v_D=\frac{4 \pi \Qn}{V} - v_*.
\label{eq:vD} 
\end{equation}
Similarly to the tracer displacement $\Delta$, the Darwin drift of a swimmer turns out independent of the quadrupole strength $\Qz$.
To appreciate the result (\ref{eq:vD}), the distinction between the volume $v_*$ and the swimmer volume $v_s$ should be emphasized: 
\begin{equation}
 v_*=v_s+v_{wake},
\end{equation}
where $v_{wake}$ is the volume of the swimmer wake.

Eq.~(\ref{eq:vD})  provides a decomposition of $v_D$ into a universal and a swimmer-dependent part. For an 
arbitrary force-free swimmer the near-field details affect the entrained volume by virtue of $v_*$ only. In the next section we illustrate the utility of the expression (\ref{eq:vD}) for two model swimmers: the squirmer model and a numerical model of the bacterial cell.

For the special case of $\Qz=\Qn=Q$, the expression (\ref{eq:vD}) reduces to   
\begin{equation}
v_D=\frac{4 \pi Q}{V} - v_*.
\label{eq:Darwin} 
\end{equation}
This formula is, in fact, identical to the one derived by \citet{Taylor} in the context of inviscid flows. But the underlying physical reasons are, of course, different. In the inviscid case the potential dipole (coinciding with the viscous force quadrupole \citep{Chwang}) is the only dominant fluid velocity component present that may contribute to $A$. In the (highly) viscous case, however, the Stokeslet and stresslet components do not contribute because of the force-free nature of the swimmer motion and the symmetry conditions on the swimmer trajectory respectively. We recently found that the expression (\ref{eq:Darwin}) was also stated in the context of swimmers in \citet{LeshanskyPismen} but the questions of universality and the role of the wake were not discussed.

We have assumed so far that microswimmers move in a roughly steady fashion and produce stationary flows. But some microswimmers, such as {\it C. rheinhardtii}, produce oscillatory flows by beating their flagella asymmetrically during the power and recovery strokes. Nevertheless, all conclusions obtained in this chapter remain valid if we understand the physical quantities such as the fluid velocity, the streamfunction and the multipole strengths in the beat cycle-averaged sense. 

\section{Two model swimmers}

\subsection{Entrainment by a spherical squirmer}

The spherical squirmer model has been the theorists workhorse for studying swimming at low {\it Re} numbers for more than half a century \citep{Lighthill, Blake, IshiPedley06}. It assumes a spherical swimmer of radius $a$ with a prescribed tangential velocity $u_{\theta}$ on the swimmer's surface in the co-moving reference frame ($\theta$ is the polar angle measured from the swimming direction). The induced flow can be found analytically as a superposition of spherical harmonics. In the simplest nontrivial case when just the first two harmonics are nonzero:
\[
u_{\theta}= B_1 \sin \theta + B_2 \sin \theta \cos \theta,
\]
where $B_1$ and $B_2$ are constants. For a neutrally buoyant, force-free swimmer $B_1$ sets the swimming speed $V=\frac23 B_1$, and $B_2$ sets the swimmer dipole strength. The squirmer streamfunction equals
\begin{equation}
\label{eq:squirmer}
\psi=\pi \rho^2 V \left[ 
1 
+ \frac{3 \hat \beta }{2} \frac{z}{r} \left(  \left( \frac{a}{r} \right)^2 - \left( \frac{a}{r} \right)^4 \right) 
- \left( \frac{a}{r} \right)^3 
\right],
\end{equation}
where $\hat \beta=B_1/B_2$. Comparison of the latter expression with (\ref{eq:streamf}) shows that squirmer dipole strength $\kappa = \frac{3 \hat \beta}{4} V a^2$ and the quadrupole strength $\Qn=\Qz=Q = \frac{1}{2}V a^3$. Clearly, squirmers having $\hat \beta>0$ are pushers while those having $\hat \beta<0$ are pullers. For strong enough dipole strength, $\vert \hat \beta \vert >1$, squirmers develop a bubble-like wake. The dependence of the squirmer wake volume on $\hat \beta$, $v_{wake}/v_0=f(\hat \beta)$, can be calculated analytically, see Appendix C.

\begin{figure}
\begin{center}
\vspace{-1cm}
\includegraphics[height = 6cm, angle=90]{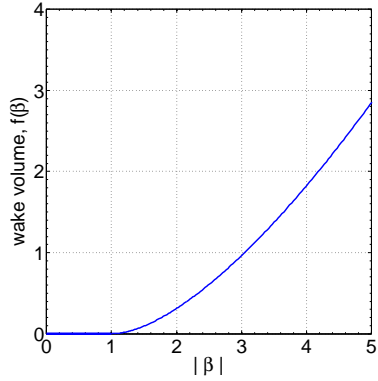}
\vspace{-1cm}
\caption{ Squirmer wake volume as a function of $\hat \beta={(4 \kappa)/ (3 V a^2)}$ where $\kappa$ is the dipole strength, $V$ the velocity and $a$ the radius of the squirmer. The wake volume is measured in units of the squirmer volume.}
\label{p:squirmwavevol}
\end{center} 
\end{figure}

Having substituted the above calculations in the expression (\ref{eq:Darwin}), we readily obtain: 
\begin{equation}
\label{eq:squirvol}
v_D = v_0 \left( 1/2 - f(\hat \beta) \right).
\end{equation}
The first term in the brackets is the added mass coefficient for a sphere $C_D=1/2$. The function $f(\hat \beta)$ is plotted on Figure \ref{p:squirmwavevol}. Thus, when $\vert \hat \beta \vert<1$ the volume entrained by a squirmer equals that of a sphere of the same size moving with the same speed in inviscid fluid.  For larger $\vert \hat \beta \vert$, $v_D$ decreases and becomes negative for $\vert \hat \beta \vert>2.4$.

\subsection{A model for R. sphaeroides}

\begin{figure}
\centering
\includegraphics[width = 0.8\textwidth]{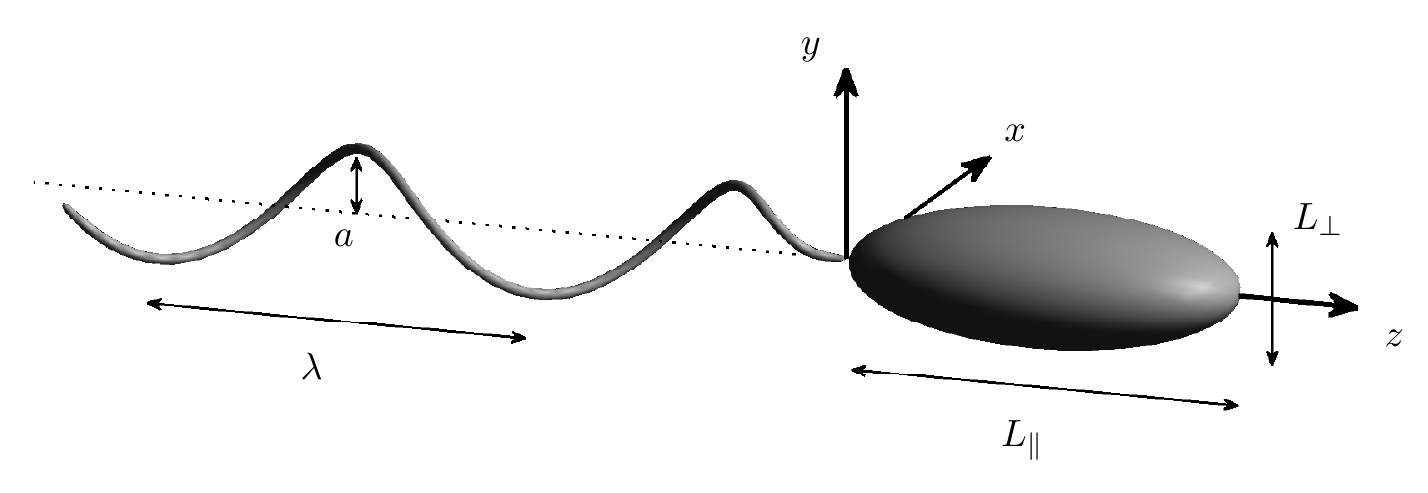}
\caption{Geometrical model of {\em Rhodobacter sphaeroides}, after~\citet{shum_modelling_2010}.}
\label{fig:bact_model}
\end{figure}

We now examine numerically the paths of passive tracer particles in the flow caused by a more realistic model swimmer, designed to represent the bacterium, {\em Rhodobacter sphaeroides} \citep{shum_modelling_2010}. The geometrical model consists of a prolate spheroidal cell body and helically-shaped flagellum of finite thickness, as depicted in Fig.~\ref{fig:bact_model}. Each of these components is rigid and the flagellum rotates about its axis, which coincides with the major axis of the cell body.

Employing the boundary element method described in \citet{shum_modelling_2010}, we numerically determine the flow field and swimming velocity of the swimmer which  satisfies the Stokes equations, no-slip boundary conditions on the surface of the cell body and flagellum, and the constraints that zero net force and torque act on the swimmer.

We begin with simulations of a bacterial swimmer with a cell body of length/width aspect ratio $L_\parallel/L_\perp = 2$ and a flagellum of helical pitch $\lambda = 2$ and length $L=10$. The non-dimensional lengthscales and timescales used throughout this section correspond respectively to $1 \mu m$ and 0.1\,s for typical bacteria. 
Using the average flow field, we compute the trajectories of tracer particles initially arranged along the line segment $z  = 0$, $0.1 \leq \rho \leq 30$. The swimmer starts at $z=-1000$ and we run the simulations until it reaches $z=1000$.  The results are extrapolated in both directions to give the tracer displacements for an infinite swimmer path (see Appendix D for details). We find that we can neglect the net radial displacement which is orders of magnitude smaller than typical displacements in the $z$-direction.

\begin{figure}
\centering
\includegraphics[height=5cm]{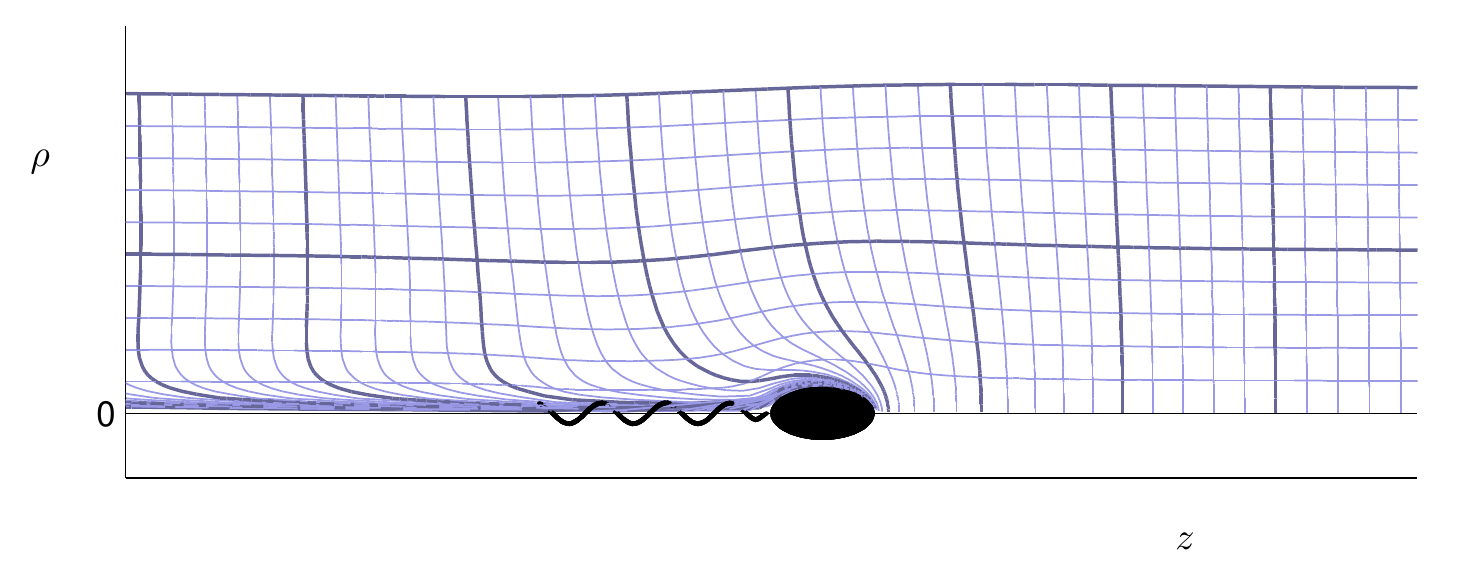}
\vspace{5mm}
\includegraphics[height=6cm]{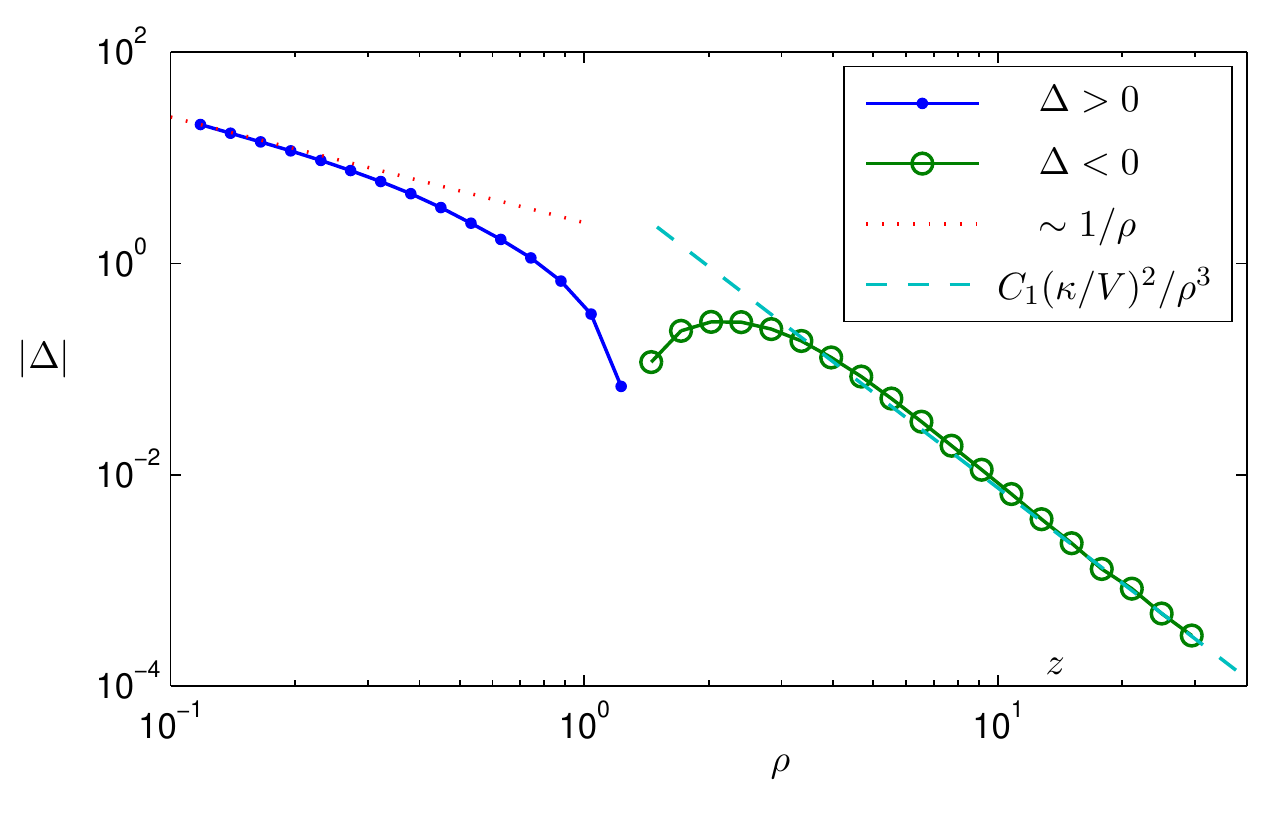}
\caption{ (a) Deformation of an initially uniform rectangular grid of tracer particles as a bacterial swimmer approaches from infinitely far away. (b) Deformation (averaged over angles) of the material sheet. Plots of $z$-component of the net displacement as functions of initial radial position on logarithmic axes. Dotted and dashed lines are guides indicating the slopes of power law scaling. }
\label{fig:tracerframes}
\end{figure}

Paths of tracer particles for different initial values of $\rho$ are shown in Fig.~\ref{fig:loops}. 
The disturbance of the fluid can also be visualised as in Fig.~\ref{fig:tracerframes}(a), which illustrates the deformation of an initially uniform rectangular grid of tracer particles as a bacterial swimmer approaches from infinitely far away. 

Double logarithmic plots of the net tracer displacements for an infinite swimmer path as functions of the initial tracer radial distance $\rho$ are presented in Fig.~\ref{fig:tracerframes}(b).  For tracers that start far from the swimmer the net displacement is negative and scales as $\rho^{-3}$ in agreement with Eq.~(\ref{eq:delta}).  For those that start near to the swimmer the net displacement is positive, and diverges as $\rho^{-1}$ as  $\rho \rightarrow 0$. This is because a fluid particle ahead and precisely on the axis of the swimmer approaches the stagnation point at the leading pole of the cell body and is pushed indefinitely by the swimmer.
 
We calculate the Darwin drift $v_D$ by a direct integration of the volume swept out by the material sheet for an infinite swimmer path, obtaining  $v_D = -6.29$ in units of cubic swimmer body radii. Details of the numerical integration are given in Appendix D. To make contact with the expression~(\ref{eq:Darwin}), we first note that the swimmer had no discernable wake. Hence $v_* = v_s = 4\pi/3\approx 4.19$. $Q_\perp$, can be obtained by numerical integration of a  moment of the force on the swimmer over the swimmer surface,
\begin{equation}
Q_\perp=-\frac12 \int_S f_z \rho^2 dS
\label{e:Qperp}
\end{equation}
giving $Q_\perp/V=0.15$. Putting these numbers into Eq.~(\ref{eq:Darwin}), gives $v_D=-6.10$, in good agreement with the value obtained by direct numerical integration.

These calculations were repeated using the flow fields generated by swimmers with a range of shapes, but keeping the volume of the swimmer body constant. In particular, we started with the base parameter values, ($L_\parallel/L_\perp = 2, \lambda = 2, L=10)$, and varied each parameter in turn using the values:
\begin{equation*}
 L_\parallel/L_\perp = 0.5, 3.5, \qquad \lambda = 0.5, 3.5, \qquad L = 5, 15.
\end{equation*}
None of the swimmers had a wake and hence $v_*=4.19$ as before. The value of  $Q_\perp/V$ for each of the geometries is listed in Table~\ref{tab:bact_volumes}. The Table then compares the values of $v_D$ calculated by using   $Q_\perp/V$ in Eq~(\ref{eq:Darwin}) to that obtained by direct numerical integration of the volume of fluid displaced, showing pleasing agreement. The results indicate that for plausible swimmer shapes, the drift is of the order of the swimmer volume. It can be seen from Eq.~(\ref{eq:vD}) and Eq.~(\ref{e:Qperp}) that the drift depends on the radial distributions of the propulsive forces on the flagellum and drag forces on the swimmer. The drift is enhanced by increasing the difference between the widths of the cell body and the flagellum helix. Swimmer shape is therefore critical in determining the Darwin drift.

\begin{table}
 \centering
\caption{Comparison of Darwin drift for different swimmer geometries. Unless listed in the first column, parameter values used are: $L_\parallel/L_\perp = 2,\, \lambda = 2,\, L=10,\, a = \lambda/2\pi$.}.
\vspace{2mm}
\begin{tabular}{cccc}
 \hline
\\[-2.2ex]
Shape       &~~~~$Q_\perp/V~~~~$ & $v_D$  & $ v_D$ \\
&&(from Eq.~(\ref{eq:vD})) &(from simulations)\\
\hline
Base                       &-0.15& -6.10 & -6.11 \\
$L_\parallel/L_\perp = 0.5$  &-0.68&-12.74 & -12.78 \\
$L_\parallel/L_\perp = 3.5$  &-0.08& -5.24 & -5.33 \\
$L = 5$                      &-0.17& -6.36 & -6.30 \\
$L = 15$                     &-0.14& -5.98 & -6.03 \\
$\lambda=0.5$              &-0.20& -6.76 & -6.78 \\
$\lambda=3.5$              &-0.04& -4.71 & -4.68 \\
$\lambda=8, L = 20$        & 0.58&  3.04 &  3.07 \\
\hline
\end{tabular}
\label{tab:bact_volumes}
\end{table}

\section{Discussion}

The aim of our paper has been  to describe the entrainment of tracer particles by microswimmers. This is a step towards a full understanding of the ways in which bacteria can stir their environment,  its likely importance in enhanced nutrient uptake and its possible relevance to oceanic mixing. Our results hold in the limit of zero Reynolds number, which is appropriate at the length scales of bacteria, and rely  heavily on the observation that bacteria are force-free so that their far field is, to leading order, dipolar.

The trajectories of tracer particles far from a swimmer that moves in a straight line along an infinite path  form closed loops. This observation is at odds with experiments and simulations showing enhanced tracer diffusion in bacterial suspensions. Therefore we first present a full derivation of the tracer displacement which allows us to list the conditions necessary for closed loop tracer trajectories. We find that the loops are closed only if the tracer velocity can be neglected compared to the swimmer velocity. As the initial distance between the tracer and the path of the swimmer $\rho$ decreases this condition is violated to an increasing degree, and a far field expansion shows that the net tracer displacement is negative and increases as  $\rho^{-3}$ for a dipolar swimmer, and as  $\rho^{-5}$ for a quadrupolar swimmer.

Numerical results for a model swimmer show that as $\rho$ decreases further the net tracer displacement changes sign. For tracers very close to the swimmer it becomes large and positive and diverges as $\rho \rightarrow 0$. This is expected as it reflects the behaviour of a tracer ahead and precisely on the axis of the swimmer that approaches the stagnation point at the leading pole of the cell body and is pushed indefinitely by the swimmer. However the total volume of fluid transferred by the passing swimmer, a quantity known as the Darwin drift, remains finite. We calculate the Darwin drift and show that it can be written simply as a difference of two terms, see Eq.~(\ref{eq:Darwin}). The first of these is, perhaps surprisingly, universal, as it depends only on the quadrupole moment and velocity of the swimmer. The second term has a pleasing physical interpretation, being equal to the sum of the volumes of the swimmer and of its wake. We show that the formula holds for squirmers, where the Darwin drift can be calculated analytically, and agrees with numerical results for a model swimmer resembling {\em Rhodobacter sphaeroides}.

A second mechanism which leads to non-closed tracer trajectories, and hence tracer diffusion, is finite swimmer run lengths. 
Mixing enhancement due to random re-orientation of straight runs was first suggested by \citet{LinChildress}. Many motile bacteria, such as  {\it E. coli} \citep{Berg} and {\it C. reinhardtii} \citep{TwoGears} have an in-build mechanism of flagellar desynchronisation that leads to a run-and-tumble manner of propulsion. It has been hypothesised \citep{Stocker09} that this feature has evolved as a stealthy strategy to avoid predators. The enhancement of biomixing may serve as an alternative evolutionary hypothesis. 

As a bacterial suspension becomes more concentrated, tracers will interact  simultaneously with several swimmers, and this will be a factor in determining their diffusivity. Moreover, in this article we have neglected Brownian motion. On average Brownian fluctuations do not destroy the loop-like topology of the tracer paths and a first approximation may be to add the Brownian and swimmer contributions to the tracer diffusion. However this remains to be verified.

To summarise, we have given an account of the effect of entrainment on the paths of tracers moving near a low Reynolds number swimmer. This is one of the mechanisms, several of which are identified here, which may lead to enhanced tracer diffusivity in swimmer suspensions. We hope that our results are a step towards determining the relative importance of various stirring mechanisms  in different regimes of swimmer concentration and for diverse swimmer strokes and trajectories.\newline

JMY and DOP acknowledge funding from the ERC Advanced Grant MiCE. DOP is grateful to Igor Kuznetsov for discussions and assistance 
with figures. 

\appendix

\section{The scaling law $\Delta \sim \rho^{-3}$}

In this section we derive the scaling law for the passive tracer displacement as a single swimmer travels in a straight infinite path. We will assume the origin at the initial tracer location $\mb r_T \big|_i= \mb 0$ and direct the z-axis along the swimmers's path, see Fig.~\ref{fig:loops} .

When the particle displacement is much smaller than the swimmer displacement, $\vert \mathbf r_T \vert \ll  \vert \mathbf R \vert$, the right hand side of the equation for the tracer motion (\ref{eq:tracer}) can be expanded as:
\begin{equation}
\frac{d \mathbf r_T}{dt}=\mathbf u(\mb R(t)) 
- \lbr \mb r_T \cdot \nabla \rbr 
\mb u(\mb R(t)) + 
O \lbr \lbr \vl \mb r_T \vl / \vl \mb R \vl \rbr^2 \rbr.
\end{equation}

We solve the above equation using the perturbation theory. For the zeroth order term $\mb r_T^{(0)}$:
\begin{equation}
\label{eq:u0}
\frac{d \mb r_T^{(0)}}{dt}=\mathbf u(\mb R(t)). 
\end{equation}

If the z-axis is directed along the swimmer's path,  the equation (\ref{eq:ufar}) can be written as:
\begin{align}
\mb u (\mb R) =  \pdz \Biggl[ &
-\kappa \mb U_S \left(  \mathbf{R} ,\mathbf{k} \right) 
\Biggr. \nonumber\\ & \Biggl. 
+ \Qn \nabla \Bigl( \frac{1}{R}  \Bigr) 
+ \frac{\Qn-\Qz}{2} 
\pdz
\mathbf{U}_S  \left(  \mathbf{R} ,\mathbf{k} \right)
\Biggr]
+ O( R^{-4}),
\end{align}
where $R=\vl \mb R \vl$ and $\pd_z=\lbr \mb k \cdot \nabla \rbr$.
Taking into account that $\frac{d}{dt}=V \pd_z$ we can integrate (\ref{eq:u0}) along the swimmer location $z$ and obtain:
\begin{align}
\mb r_T^{(0)}(z)&=\lsb - \frac{\kappa}{V}\mb{U_s}(\mb R(z),\mb k) \right. \nonumber\\
& \left.
+ \frac{\Qn}{V} \nabla \frac{1}{R} 
+ \frac{\Qn-\Qz}{2V} 
\pdz
\mb{U_s}(\mb R(z),\mb k)
\rsb_{\mb R(z)} + O(R(z)^{-3}).
\end{align}
In determining the integration constant in the above expression we used the initial conditions $z_i=-\infty$ and $\mb r_T^{(0)} \big|_i= \mb 0$. It can be immediately seen that as the swimmer goes to infinity $z \to +\infty$, $\vl \mb r_T^{(0)}\vl \sim R(z)^{-1} \to 0$,  i.e. the tracer trajectory is a closed loop, in agreement with the general result formulated in Section 2. Therefore the tracer displacement will be dominated by the first order perturbation $\Delta \mb r_T^{(1)}$.

The equation for $\mb r_T^{(1)}$ reads:
\begin{equation}
\frac{d \mb r_T^{(1)}}{dt}= - \lbr \mb r_T^{(0)} \cdot \nabla \rbr 
\mb u(\mb R(t)).
\end{equation}
Substituting in the expressions for $\mb u$ and $\mb r_T^{(0)}$ obtained above and keeping the terms up to the order $O(R^{-5})$,
we obtain:
\begin{align}
\frac{d \mb r_T^{(1)}}{dt} = & -\frac{\kappa^2}{V} \lbr \mb U_S \cdot \nabla \rbr \pd_z \mb U_S 
 \nonumber\\ & 
+ \frac{\kappa \Qn}{V} \lsb
\lbr \mb U_S \cdot \nabla \rbr \pd_z \nabla \frac{1}{R} 
+ \lbr \nabla \frac{1}{R} \cdot \nabla \rbr \pdz \mb U_S 
\rsb  \nonumber\\ & 
+ \frac{\kappa (\Qn-\Qz)}{2 V} \pdz \lsb \lbr \mb U_S \cdot \nabla \rbr \pdz \mb U_S \rsb
+O(R^{-6}).
\label{e:u1}
\end{align}
The latter equation can be integrated along $dz=V dt$. The term with the prefactor proportional to $(\Qn-\Qz)$ is a derivative along the path and will, therefore, vanish upon integration from $-\infty$ to $+\infty$. It is easy to see that the result has to be of the form:
\begin{equation}
\Delta \mb r_T^{(1)}=-\mb C_1 \frac{\kappa^2}{V^2} \frac{1}{\rho^3}
+\mb C_2 \frac{\kappa \Qn}{V^2} \frac{1}{\rho^4}
+O \lbr \rho^{-5}  \rbr,
\end{equation}
where $\rho$ is the distance between the swimmer's path and the initial position of the tracer, and $\mb C_1, \; \mb C_2$ are (vector) constants. It turns out that the displacement in the direction normal to the swimming direction is negligible when the swimmer's path is infinite. Therefore we will focus on the tracer displacement along the swimming direction $\Delta=\lbr \mb k \cdot \Delta \mb r_T \rbr$:
\begin{equation}
\Delta = - C_1 \frac{\kappa^2}{V^2} \frac{1}{\rho^3}
+ C_2 \frac{\kappa Q}{V^2} \frac{1}{\rho^4}
+O \lbr \rho^{-5}  \rbr.
\label{e:asym}
\end{equation}
Here $C_1$ and $C_2$ are scalars.
Let us calculate the numerical value of $C_1$.
According to (\ref{e:u1}) and (\ref{e:asym}),
\begin{equation}
C_1=\int_{-\infty}^{+\infty} dz \rho^3 \lbr \mb U_S \cdot \nabla \rbr 
\pd_z \lbr \mb k  \cdot \mb U_S \rbr.
\end{equation}
Detailed calculations show that $C_1=\pi/16 \approx 0.2$.
We note in conclusion that when $\kappa=0$, the dominant contribution to $\Delta$ scales as $\rho^{-5}$.

\section{The Darwin drift}
\subsection{Streamfunction}

For a cylindrically symmetric flow the streamfunction $\psi(\rho,z)$, defined as 
\begin{equation}
\psi(\rho,z)=\int -d\rho 2\pi \rho v_z + dz 2 \pi \rho v_{\rho},
\label{e:psi}
\end{equation}
has a clear physical meaning: it equals the volume flux of incompressible fluid crossing a surface of revolution obtained by rotating  about the z-axis a two dimensional curve in the coordinates $(\rho, z)$. Without loss of generality we will assume that one of the curve ends lies on the z-axis and $\psi(\rho=0,z)=0$. Incompressibility  of the flow guarantees that $\psi$ is a function of the coordinates $(\rho, z)$ independent of particular curve choice. According to our definitions, for a uniform flow $u_\rho=0, \; u_z=-V$ and the streamfunction 
\begin{equation}
\frac{1}{2 \pi} \psi_0(\rho,z)=\frac12 V\rho^2.
\label{e:psi0}
\end{equation}

In order to find the asymptotic shape of the streamfunction for a force-free swimmer, we note that in the frame co-moving with the swimmer with the speed $V$, $v_z=-V+u_z$. Then, according to (\ref{e:psi}),
\[
\frac{\pd \psi}{ \pd \rho} = - 2 \pi \rho v_z=  2 \pi \rho (V - u_z).
\]
Therefore,
\begin{equation}
\frac{\psi}{2 \pi}= \frac12 V \rho^2 - \sideset{}{'} \int d\rho \rho u_z,
\label{e:psif}
\end{equation}
where the prime denotes that the integral is taken along $z=\text{const}$. 
According to (\ref{eq:ufar}), the flow generated by the swimmer
\begin{align}
&u_z= - \lb \kappa \pdz + \frac{\Qz}{2} \pdz^2 +\frac{\Qn}{2} \pdn^2 \rb U_{Sz}
 + O  \left(  r^{-4} \right), \nonumber\\
&U_{Sz} = \lbr \frac{1}{r} + \frac{z^2}{r^3}, \rbr \nonumber
\end{align}
where $r$ is the distance from the swimmer.
Then is it straightforward to calculate that
\begin{align}
& \pdz U_{Sz} = r^{-2} \lbr (z/r)-3(z/r)^3 \rbr, \\
& \pdz^2 U_{Sz} = r^{-3} \lbr 1 - 12 (z/r)^2 + 15 (z/r)^4 \rbr, \\
& \pdn^2 U_{Sz} = r^{-3} \lbr 1 + 6 (z/r)^2 - 15 (z/r)^4 \rbr.
\end{align}

We obtain upon integration and a few straightforward transformations
\begin{align}
- \sideset{}{'} \int d\rho \rho u_z & = \,
\kappa \frac{\rho^2 z}{r^3} 
 \nonumber\\ &
- \frac{\rho^2}{r^3}
\left( 
 \Qz \frac{1-3 \cos^2 \theta}{2} 
+ \Qn \frac{1+3 \cos^2 \theta}{2} 
\right) \nonumber\\ &
+ O(r^{-2})+C(z).
\end{align}
Here $C(z)$ is an arbitrary function at this point. In order to determine it we recall
\[
\frac{\pd \psi}{ \pd z} =  2 \pi \rho v_\rho=  2 \pi \rho u_\rho.
\]
It can be verified that this expression holds if $C'(z)=O(z^{-4})$ for large $z$. Therefore $C(z)=C_1+O(z^{-3})$. As far from the swimmer its influence on the flow must be negligible, we conclude that $C_1=0$ and arrive at the final expression:
\begin{align}
\frac{\psi}{2 \pi} =  \rho^2 & \lsb  \frac{V}{2} + \frac{\kappa z}{r^3} \right. \nonumber \\
 & \left.   - \lb 
\Qz \frac{1-3 \cos^2 \theta}{2} + \Qn \frac{1+3 \cos^2 \theta}{2} 
\rb 
\frac{1}{r^3} \rsb  + O(r^{-2}). \nonumber 
\end{align}

\subsection{Drift}

Following the original argument of \citet{Darwin}, we consider a circular material sheet of radius $\rho_0$ placed in front of the swimmer at the initial distance $L_i$, Figure~\ref{DarwinPic}. In the co-moving reference frame the incoming flow is uniform and has the speed $V$. If the swimmer did not perturb the flow, the material sheet would travel with the flow and after a time $\Delta t$ assume the position $z=L_f=L_i-V \Delta t$. The fluid volume enclosed within the region swept by the material sheet, equals the total fluid volume through the boundary of this region during $\Delta t$, i.e. it equals $\psi_0(\rho_0,L_i) \Delta t$, where $\psi_0$ is the unperturbed flow streamfunction.

When the incoming flow is perturbed by the swimmer, the total amount of fluid swept by the material sheet equals $\psi(\rho_0,L_i) \Delta t$. It can be seen from Figure~\ref{DarwinPic} that 
\begin{equation}
\psi \Delta t + v_D + v_* - A = \psi_0 \Delta t
\end{equation}
where $v_D$ is the total volume of ambient fluid entrained by the swimmer, $v_*$ is the fluid volume enclosed by a stagnation streamline around the swimmer,  and $A$ is the ``excess" volume equal to
\begin{equation}
A(\rho_0) = \int_{L_f}^{L_i} dz \, \pi (\rho^2(z)-\rho_0^2),
\label{e:defA}
\end{equation}
where $\rho(z)$ is the streamtube width. In the limit of large $\rho_0$ and $L_i$, $\psi \to \psi_0$ and we obtain
\begin{equation}
v_D = A(\rho_0) - v_*.
\label{e:A}
\end{equation}

Let us find the asymptotic shape of $A(\rho_0)$. 
We re-write the streamfunction (\ref{eq:streamf}) as:
\begin{equation}
\label{eq:streamf2}
 \frac{\psi}{\pi V} =  \rho^2 \left( 1 + \text{B}(\theta) \frac{a^2 }{\rho^2} 
-  \Gamma (\theta) \frac{a^3}{\rho^3} \right)  + O((a/\rho)^{-2}),
\end{equation}
where $\text{B}( \theta) =\beta \sin^2 \theta \cos \theta$, $\Gamma (\theta)= \sin^3 \theta \lbr \gammaz \frac{1-3 \cos^2 \theta }{2} 
+ \gamman \frac{1+3 \cos^2 \theta }{2}  \rbr$, $\beta=2 \kappa / (V a^2)$, $\gammaz=2 \Qz /(V a^3)$ and $\gamman=2 \Qn /(V a^3)$.
Far from the body $\psi \to \psi_0=\pi V \rho^2$, therefore on the streamline starting at $\rho=\rho_0, \; z=\infty$,
\begin{equation}
\frac{\psi(\rho,z)}{\pi V}=\rho_0^2.
\label{e:rho02}
\end{equation}
Substituting (\ref{e:rho02}) in (\ref{eq:streamf2}) and denoting $x=a/\rho_0, \; y=\rho/\rho_0$, results in
\[
y^2 (1+ \text{B} x^2 y^{-2} -\Gamma x^3 y^{-3}+ O(\rho_0^{-4})) =1, \quad \rho_0 \to \infty.
\]
After resolving this expression with respect to $y$ we obtain
\[
y=1-\frac12 \text{B} x^2+\frac12 \Gamma x^3 + O(x^4), \quad x \to 0.
\]
We are now ready to calculate the integral in (\ref{e:defA}) in the limit of large $\rho_0$. 
\begin{align}
A&=\pi \rho_0^3 \int_{i}^{f} d \lbr y \cot \theta  \rbr (y^2-1)\nonumber \\
&=\pi \rho_0^3 \int_{i}^{f} d \lsb ( 1-\frac12 \text{B} x^2+\frac12 \Gamma x^3 + O(x^4) ) \cot \theta  \rsb \lbr -\text{B} x^2 + \Gamma x^3 + O(x^4) \rbr \nonumber \\
&=\pi \rho_0^3 \int_{i}^{f} d \lbr \cot \theta  \rbr   \lbr -\text{B} x^2 + \Gamma x^3 + O(x^4) \rbr \nonumber \\
&=\pi \rho_0^3 \int_{i}^{f} d  \theta   \lsb \beta \cos \theta x^2 - \sin \theta \lbr \gammaz \frac{1-3 \cos^2 \theta }{2} 
+ \gamman \frac{1+3 \cos^2 \theta }{2}  \rbr x^3 + O(x^4) \rsb \nonumber \\
&=\left[ \pi \beta a^2 \rho_0 \sin \theta + \pi \frac{\gammaz+\gamman}{2} a^3 \cos \theta 
+ \pi \frac{\gamman-\gammaz}{2} a^3 \cos^3 \theta
\right]_i^f \nonumber + O(\rho_0^{-1}) \\
&= \frac{2 \pi \kappa}{V} \rho_0 \sin \theta \bigg|_i^f
+ \frac{ \pi (\Qz+\Qn)}{V} \cos \theta 
+ \frac{ \pi (\Qn-\Qz)}{V} \cos^3 \theta 
\bigg|_i^f + O(\rho_0^{-1}).
\end{align}

\section{A squirmer wake}

We derive the expression for $f(\beta)$. For a squirmer
\[
\frac{\psi}{\pi V} = \rho^2 \lsb 1 +\frac{3 \hat \beta}{2} \lbr
\lbr \frac{a}{r} \rbr^2 - \lbr \frac{a}{r} \rbr^4 \rbr \cos \theta 
 - \frac{a}{r}^3  \rsb.
\]
On the stagnation streamline $\psi=0$. This occurs when 
\begin{equation}
r=a,
\end{equation}
i.e. on the squirmer surface, or on the surface
\begin{equation}
- \cos \theta = \frac{r_1 (r_1^2+r_1+1)}{(3\hat \beta/2)
(r_1+1)
} , \quad r_1=r/a .
\label{e:theta}
\end{equation}
This surface intersects the swimmer body at an angle
\[
\cos \theta_*=-1/\hat \beta.
\]
Thus, the wake exists only if $\vl \hat \beta \vl > 1.$ Let us calculate its volume.
\begin{align}
v_{wake}&=\int_{wake} dr \, d\theta \, 2 \pi r^2 \sin \theta = \frac43 \pi a^3 f(\hat \beta),\\
f(\hat \beta)&=\frac32 \int_{1}^{r_*(\hat \beta)} dr_1 \, r_1^2 (1 + \cos \theta(r_1)),
\label{e:intf}
\end{align}
where $\theta(r_1)$ satisfies (\ref{e:theta}) and $r_*(\hat \beta)$ satisfies
\[
r_*(r_*^2+r_*+1)-(3 \hat \beta/2)(r_*+1)=0.
\]
The integral in (\ref{e:intf}) can be taken analytically:
\[
f=\frac{r_*^3-1}{2}+{ \hat \beta}^{-1} \lsb 
\frac{31}{30} 
+ \log \lbr \frac{r_*+1}{2} \rbr -r_* + \frac{r_*^2}{2} -\frac{r_*^3}{3}-\frac{r_*^5}{5}
\rsb.
\]
The dependence $f(\hat \beta)$ is shown in Figure \ref{p:squirmwavevol}.

\section{{\em R. sphaeroides} simulation details}

\noindent
{\em Calculating the flow field:}
As we deal only with the case of a swimmer with an axisymmetric body in unbounded fluid, we need only compute the velocities once using the boundary element method. The velocities are constant in the reference frame of the flagellum and hence, we can calculate the mean direction and speed of propagation~\citep{keller_swimming_1976}. We choose the mean swimming direction to be along the $z$-axis and scale the spatial and temporal variables so that the volume of the cell body is $4\pi/3$ cubic units, the volumetric radius is 1, and the mean swimming speed is 1.

With typical geometrical parameters for the bacterial shape, we find that as the flagellum turns, the direction of the major axis of the cell precesses with a negligible amplitude about the direction of mean swimming. We therefore make the approximation that the swimmer is orientated in the $z$-direction at all times. We find that the period of the turning flagellum is relatively short; more than 10 complete turns are required for the swimmer to advance by one body length. We therefore consider the motion averaged over a period of revolution.

We tabulate the flow velocity on a grid of $\sim$\,35,000 points.
The grid spans the ranges $-1000\leq z \leq 1000$ and $0 \leq \rho \leq 150$, and a higher density of points is used near the swimmer to more accurately capture the variations in velocity. Between grid points, we use a cubic spline interpolation to approximate the average flow field.

To characterise the effect of a bacterial swimmer on the surrounding medium, we track a number of points in the fluid that move with the local average flow field,  using the MATLAB ODE solver, \texttt{ode113} to integrate Eq.~(\ref{eq:tracer}) in time.

\noindent
{\em Extrapolation to an infinite swimmer path}:
We begin simulations at time $t = -1000$, which means the swimmer is at position $z=-1000$ and the trajectory of each tracer particle is computed until the swimmer passes the particle and reaches the distance $z = 1000$.
To extrapolate the net tracer displacements to an infinite swimmer path we note that, for a dipolar far field flow, the tracer positions will follow the limiting behaviour
\begin{equation}
z (t) \sim \begin{cases} z (-\infty) + C^z_{-\infty}/t, & t \rightarrow -\infty, \\
                        z(\infty) + C^z_{\infty}/t, & t \rightarrow \infty. \end{cases}
\end{equation}
A nonlinear least squares algorithm is then used to estimate the limiting positions $z (-\infty)$ and $z (\infty)$, and hence the net displacement along the $z$-direction, $\Delta = z(\infty) - z(-\infty)$, for an infinite swimmer path from the finite computed trajectories.

\noindent
{\em Calculating the entrained volume}:
Net tracer displacements are obtained numerically for initial radial tracer positions $0.1<\rho < 30$. The curve is extrapolated to all values of $\rho$ using the ansatz
\begin{equation}
\Delta(\rho) = \begin{cases} C_0/\rho, & 0 < \rho  < 0.1, \\ C_\infty/\rho^3, & 30 < \rho  < \infty, \end{cases}
\label{eq:xrho}
\end{equation}
(see Fig.~\ref{fig:tracerframes}(b)).
The curve of net tracer displacements $\Delta$ can be divided into a negative region and a positive region with the transition at the point $\rho  = \rho_0$, say.
Obtaining $\rho_0$  by interpolation of the numerical data, 
we calculate the volumes of the bodies of revolution in the two regions of positive and negative net tracer displacements by evaluating the formulas
\begin{equation}
v_0 = \int_0^{\rho_0} 2\pi \rho \Delta (\rho) \; d\rho, \quad v_\infty = \int_{\rho_0}^\infty 2\pi \rho \Delta (\rho) \; d\rho,
\end{equation}
using quadrature within the range $0.1 < \rho < 30$ and analytical integration of (\ref{eq:xrho}) for the contributions outside this range. The numerical results show that typically, a volume of fluid about twice that of the swimmer is pulled forward with the swimmer across any given interface perpendicular to the swimming direction. At the same time, fluid further away from the path of the swimmer is pushed backwards. The net flux is the Darwin drift, given for different swimmer shapes in Table~\ref{tab:bact_volumes}.

\bibliographystyle{jfm}


\end{document}